\documentclass[pra,twocolumn,superscriptaddress]{revtex4-2}
 \usepackage{float,graphicx,multirow, amsmath,color,dsfont}
 \usepackage{amsmath,bm}
 \usepackage{amssymb,mathrsfs,dsfont}
 \usepackage{amssymb,mathbbol}
 \usepackage{amsfonts}
 \usepackage{mathrsfs}
 \usepackage{float}
 \usepackage{xcolor,hyperref}
 \definecolor{darkblue}{rgb}{0,0.0.1,0.3}
 \definecolor{darkred}{rgb}{0.6,0.1,0}
 \hypersetup{colorlinks,breaklinks,
 	linkcolor=darkred,urlcolor=darkblue,
 	anchorcolor=darkred,citecolor=
 	darkred,pdfauthor=ChSaAr, pdftitle=ChSaAr.Optimization}
 \usepackage{tikz}
 \usepackage{mathbbol}
 \usepackage{float}
 \usepackage[normalem]{ulem} 

 \newcommand{\ie}{\textit{i}.\textit{e}.}
 
\begin{document}
 	\title{Optimization of state parameters in displacement assisted
 		photon subtracted measurement-device-independent quantum key distribution}
 
 	\author{Chandan Kumar}
 	\email{chandan.quantum@gmail.com}
 	\affiliation{Department of Physical Sciences,
 		Indian
 		Institute of Science Education and
 		Research Mohali, Sector 81 SAS Nagar,
 		Punjab 140306 India.}
 			\author{Sarbani Chatterjee}
 		\email{mp18015@iisermohali.ac.in}
 		\affiliation{Department of Physical Sciences,
 			Indian
 			Institute of Science Education and
 			Research Mohali, Sector 81 SAS Nagar,
 			Punjab 140306 India.}
 	\author{Arvind}
 	\email{arvind@iisermohali.ac.in}
 	\affiliation{Department of Physical Sciences,
 		Indian
 		Institute of Science Education  and
 		Research Mohali, Sector 81 SAS Nagar,
 		Punjab 140306 India.}
 	\begin{abstract}
 		Non-Gaussian operations, in
 		particular, photon subtraction (PS), have been
 		shown to enhance the performance of various quantum
 		information processing tasks including continuous variable measurement device independent quantum key
 		distribution (CV-MDI-QKD). This work investigates
 		the role of non-Gaussian resource states, namely, the photon
 		subtracted two-mode squeezed coherent (PSTMSC) (which include 
 		photon subtracted two-mode squeezed vacuum (PSTMSV) as a
 		special case) states in CV-MDI-QKD.  To this end, we
 		derive the Wigner characteristic function for the resource
 		states, from which the covariance matrix and, finally, the
 		secret key rate expressions are extracted. The optimization of the state parameters is undertaken to find the most
 		suitable resource states in this family of states.
 		There have been previous studies on the PSTMSV and PSTMSC states in
 		CV-MDI-QKD that make use of PS operation.   We
 		evaluate such proposals and find to our surprise
 		that both PSTMSC and PSTMSV
 		resource states underperform as compared to the TMSV state
 		rendering PS operation   and displacement
 		undesirable.
 	\end{abstract}
 	\maketitle

 	\section{Introduction}
 	Quantum key distribution (QKD)~\cite{gisin-rep-2002,
 		scarani-rmp-2009,Pirandola-2019} is one of the most
 	developed applications of quantum information science, where
 	two distant parties, Alice and Bob, desire to establish a
 	shared secret key. The field of QKD can be broadly divided
 	into two major categories, discrete variable (DV)
 	QKD~\cite{Bennett-84, Ekert-prl-1991, Bennett-1992,
 		Wang2013} and continuous
 	variable (CV) QKD~\cite{Ralph-pra-1999,
 		Hillery-pra-2000, Cerf-pra-2001, Grosshans-prl-2002,
 		Grosshans2003, Weedbrook-rmp-2012}. CV-QKD offers a higher
 	secret key rate (SKR) than DV-QKD and has gained preference
 	over DV-QKD due to ease in
 	implementation and the possibility of carrying out 
 	homodyne and heterodyne measurements with high efficiency.
 	Security aspects of
 	some of these protocols have been analyzed in detail in Refs.~\cite{Leverrier-prl-2013,
 		Leverrier-prl-2015, Leverrier-prl-2017, Renner-prl-2009,
 		Shor-prl-2000}.
 	Unconditional security of QKD protocols typically
 	requires perfect devices and noiseless channels. In the
 	practical situations where noise is always present and devices
 	are not perfect, security loopholes develop and constructing
 	a fully secure protocol becomes a challenge.

 	The DV measurement device independent (MDI) QKD protocol,
 	which is based on the concept of entanglement swapping, was
 	proposed to remove all security loopholes arising from the
 	detectors~\cite{Braunstein-prl-2012, Lo-prl-2012}. Soon
 	after, the idea of MDI was extended to CV
 	systems~\cite{Pirandola-np-2015, Li-pra-2014,
 		Xiang-pra-2014}. In CV-MDI-QKD protocols, an untrusted third
 	party employs a CV Bell measurement and publicly
 	communicates the results, which leads to  sharing of
 	the secret key between the legitimate parties. The SKR obtained
 	in CV-QKD is higher; however, the maximum transmission
 	distance  is substantially lower
 	as compared to DV-MDI-QKD.

 	Typically multimode Gaussian states of the radiation field
 	are used as resource states for general   quantum information processing (QIP)  tasks and in
 	particular for CV-MDI-QKD~\cite{Weedbrook-rmp-2012,Pirandola-2019}.
 	In order to increase non-classicality and entanglement,
 	non-Gaussian states have been engineered by implementing
 	non-Gaussian operations such as photon subtraction,
 	addition, and catalysis on Gaussian states~\cite{Agarwal-pra-1991,Kitagawa-pra-2006,Ourjoumtsev-prl-2007,Takahashi-nature-2010,Zhang-pra-2010,pspra2012}.
 	Such non-Gaussian states have been employed to improve the
 	performances of various   QIP 
 	protocols such as quantum
 	teleportation~\cite{tel2000,dellanno-2007,tel2009,catalysis15,catalysis17,wang2015,tele-2023,noisytele},
 	quantum key
 	distribution~\cite{qkd-pra-2013,Ma-pra-2018,chandan-pra-2019,qkd-pra-2019,qk2019,zubairy-pra-2020}
 	and quantum
 	metrology~\cite{gerryc-pra-2012,josab-2012,braun-pra-2014,josab-2016,pra-catalysis-2021,ill2008,ill2013,metro22,metro-thermal-arxiv, ngsvs-arxiv}.
 	These non-Gaussian operations, in particular, the photon
 	subtraction (PS) on two-mode squeezed vacuum (TMSV) and
 	two-mode squeezed coherent (TMSC) states, have been shown to
 	enhance the performance of CV-MDI-QKD
 	protocols~\cite{Ma-pra-2018,chandan-pra-2019}. 
 	
 	In this work, we undertake to explore the full
 	parameter space of PSTMSC states for their
 	performance of the CV-MDI-QKD protocol. The photon
subtracted TMSC (PSTMSC) is a family of
 	states determined by three parameters, namely, variance (squeezing),
 	displacement, and transmissivity of the beam splitter
 	involved in the implementation of the PS operation.
 	We first keep the variance fixed and
 	optimize the SKR with respect to the remaining two
 	state parameters. At fixed high variances, we obtain an
 	improvement in the maximum transmission distance and a
	significant improvement in the SKR as compared to
the results reported earlier~\cite{chandan-pra-2019}.
Further, we observe that the PSTMSV states do not
 	provide any advantage either in SKR or maximum transmission
 	distance over the TMSV state, unlike the conclusion
 	drawn in an earlier
 	work~\cite{Ma-pra-2018}.  At fixed low variance,
 	the results show that photon subtraction operation  does not
 	provide any benefit in terms of higher SKR or transmission
 	distance.  Finally, we  perform the optimization of SKR with
 	respect to all the three state parameters.  In
 	contradiction to several recent papers
 	claiming that photon subtraction and displacement on
 	Alice's side of the CV-MDI-QKD protocol is advantageous, we
 	conclude that there is no advantage in using photon
 	subtraction and displacement in CV-MDI-QKD protocols.
 	
  Our study directly impacts the utility of photon subtraction in 
 	CV-MDI-QKD protocols~\cite{virtualmdi, amplifiers,underwater}. Other 
 	CV-QKD protocols such as entanglement-based CV-QKD protocol~\cite{ebcvqkd}, 
 	entanglement in the middle CV-QKD protocol~\cite{middle}, 
 	virtual post selection based CV-QKD protocol~\cite{virtual16} also needs to be 
 	optimized with respect to state parameters to assess their utility of photon subtraction.

 	Exploiting the fact that the introduction of
 	displacement  also helps maintain the maximum transmission
 	distance for multi-photon subtracted TMSC states, we propose
 	a new variant of the QKD protocol that simultaneously uses
 	1-PSTMSC, 2-PSTMSC, 3-PSTMSC and 4-PSTMSC states
 	($k$-PSTMSC state represents $k$ photon subtracted TMSC
 	state). This enables us to maximize the utilization of
 	resources per trial and consequently leads to a significant
 	enhancement in the key rate.   We note that this protocol is efficient at high variance. Our analysis   at variance $V=15$  shows that using
 	1-PSTMSC state alone provides a resource utilization of
 	approximately $36\%$, while using all four PSTMSC states
 	enhances it to approximately $78\%$.

 	Other  important aspect of the current work is the derivation
 	of the  Wigner characteristic function of the PSTMSC states.
 	The elements of the covariance matrix,  which are
 	second-order moments of the quadrature operators symmetrized
 	as per the Weyl correspondence rule, can be evaluated by
 	differentiation of the Wigner characteristic function. The
 	Wigner characteristic function, as well as the different
 	moments of the covariance matrix, are provided in a compact
 	form, which will render its applications in other QKD
 	protocols to be convenient.
 	
 	The article is structured as follows.    Section~\ref{app:pstmsc} contains the derivation of the
 	Wigner characteristic function and the different moments of
 	the covariance matrix of the PSTMSC states.
 	In Sec.~\ref{cvmdipstmsc},
 	we briefly describe the  PSTMSC states based
 	CV-MDI-QKD protocol and in Sec.~\ref{skr}, we set up the notations required to
 	evaluate the secret key.  
 	In Sec.~\ref{sec:opt}, we optimize
 	the SKR with respect to the state parameters.  Finally, in
 	Sec.~\ref{sec:conc}, we  conclude the main results of the
 	current work and discuss future prospects.  In
 	Appendix~\ref{app:char}, we review the formalism of CV
 	systems and its phase space description.

 	\section{Review of CV-MDI-QKD protocol}
 	\label{cvmdi}

 	In this section, we   evaluate the covariance matrix required for the calculation fo secret key rate. Further, it provides a brief overview of the
 	entanglement-based CV-MDI-QKD protocol for the PSTMSC
 	state~\cite{chandan-pra-2019}   and details its security analysis.  
 	
 	\subsection{ Wigner characteristic function of the PSTMSC state}\label{app:pstmsc}
 	\begin{figure}[h!] 
 		\centering
 		\includegraphics[scale=1]{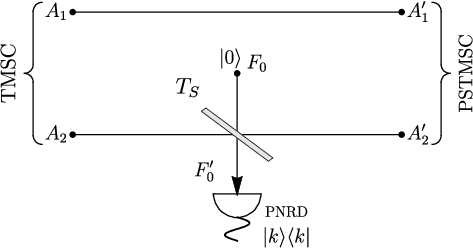}
 		\caption{Schematic for preparation of the PSTMSC state
 			generation from the TMSC state. The mode $A_2$ of the
 			TMSC state is interfered with the ancilliary mode $F_0$
 			using a beam splitter. The detection of $k$ photons at the
 			PNRD represents a successful preparation of a $k$-PSTMSC
 			state.}
 		\label{pstmsc_sub}
 	\end{figure}
 	The schematic for the generation of the PSTMSC state is
 	shown in Fig.~\ref{pstmsc_sub}. The two modes of the TMSC state are  labeled  as $A_1$ and $A_2$ and
 	represented by the quadrature operators
 	$\hat{q}_1,\hat{p}_1$ and $\hat{q}_2,\hat{p}_2$. 
 	To generate TMSC state, one begins with two uncorrelated
 	modes, each in a
 	coherent state, represented by the following
 	displacement vector and covariance matrix (in shot noise
 	units):
 	\begin{equation}
 		\bar{\hat{\xi}} = ( d,0,d,0)^T,\quad \Sigma= \mathbb{1}_4,
 		\label{disp_cov}
 	\end{equation}
 	where $\mathbb{1}_4$ is the $4 \times 4$ identity
 	matrix.  These modes
 	are sent through a two-mode nonlinear optical down converter
 	to obtain a TMSC state.   The displacement vector and
 	covariance matrix transform according to
 	Eq.~(\ref{transformation}) given in Appendix~\ref{cvsystem}
 	as
 	\begin{equation}\label{tmscmean}
 		\bar{\hat{\xi}} \rightarrow  S_{A_1 A_2 }(r)\,
 		\bar{\hat{\xi}},\quad \Sigma \rightarrow S_{A_1 A_2 }(r)  \Sigma S_{A_1 A_2
 		}(r)^T,
 	\end{equation}
 	where  $S_{A_1 A_2 }(r)$ is the two mode squeezing
 	operator defined in Eq.~(\ref{eq:tms}). 	We use the Wigner characteristic function formalism,
 	which turns out to be a convenient and effective approach
 	in calculating the covariance matrix, whose elements
 	are second order moments symmetrized as per the Weyl
 	correspondence rule.
 	The Wigner
 	characteristic function~(\ref{wigc}) of the TMSC state
 	evaluates to 
 	\begin{equation}\label{charwigtmsc}
 		\begin{aligned}
 			\chi_{A_1 A_2}&(\Lambda_1,
 			\Lambda_2)= \exp \bigg[ -\frac{1}{2} \cosh (2 r)
 			\left(\sigma _1^2+\sigma _2^2+\tau _1^2+\tau _2^2\right) \\
 			& +\sinh (2 r) \left(\tau _1 \tau
 			_2-\sigma _1 \sigma _2\right)+id
 			e^{-r}\left(\sigma_1+\sigma_2\right)\bigg].
 		\end{aligned}
 	\end{equation}
 	
 	We now consider an ancilla mode $F_0$ represented by the
 	quadrature operators $(\hat{q}_3,\hat{p}_3 )^T$ initialized
 	to the vacuum state.  Fred interferes the ancilla mode $F_0$
 	with the mode $A_2$ of the TMSC state received from Alice,
 	with the help of  a beam splitter of transmissivity $T_S$.
 	The Wigner characteristic function prior to the interference
 	is given by
 	\begin{equation}
 		\chi_{A_1 A_2F_0}(\Lambda ) = \chi_{A_1 A_2}(\Lambda_1,
 		\Lambda_2)\chi_{|0\rangle}(\Lambda_3).
 	\end{equation} 
 	The action of the beam splitter entangles the input
 	modes and the transformed Wigner characteristic function
 	post the beam splitter interference is given by
 	\begin{equation}
 		\chi_{A'_1 A'_2F'_0}(\Lambda ) = \chi_{A_1
 			A_2F_0}\left([\mathbb{1}\oplus B_{A_2 F_0}(T_S)]^{-1}\Lambda
 		\right).
 	\end{equation}
 	Now the transformed mode $F'_0$ is measured using a PNRD,
 	which is described by the POVM $\{ \Pi_k = |k\rangle \langle
 	k|, \mathbb{1}-\Pi_k \}$.
 	A click of the POVM element 
 	$\Pi_k$ represents a successful $k$ photon
 	subtraction. The unnormalized Wigner characteristic function
 	of the $k$-PSTMSC state is given by
 	\begin{equation}\label{wigred}
 		\begin{aligned}
 			\widetilde{ \chi}^{(k)}_{A_1 A'_2}
 			(\Lambda_1, \Lambda_2)& =\frac{1}{  \pi} \int d^2 \Lambda_3
 			\,\underbrace{ \chi_{A'_1 A'_2F'_0}(\Lambda )
 			}_{\text{Three mode entangled state}}\\
 			& \times \underbrace{\chi_{|k\rangle
 				}(-\Lambda_3)}_{\text{Projection on }|k\rangle \langle k|}.
 		\end{aligned}
 	\end{equation}
 	To integrate the above equation, we write the
 	Laguerre polynomial occurring in the Wigner characteristic
 	function of the Fock state $|k\rangle$  as
 	\begin{equation}
 		L_k\left(\tau_3^2+\sigma_3^2\right)
 		=\frac{1}{k!} \frac{\partial^{k}}{\partial u
 			^{k}}\frac{\partial^{k}}{\partial v^{k}} e^ {uv+u
 			(\tau_3+i\sigma_3)-v (\tau_3-i\sigma_3) } \bigg|_{u=v=0}.
 	\end{equation}
 	Substituting  the above in Eq.~(\ref{wigred}) leads to a Gaussian 
 	integral, which evaluates to
 	\begin{equation}\label{reducedcal}
 		\begin{aligned}
 			\widetilde{ \chi}^{(k)}_{A_1 A'_2}&(\Lambda_1, \Lambda_2) = 2 z_1 z_2^{k}
 			\exp\big[  x_1 \left(\sigma_1^2+\tau_1^2+\sigma_2^2+\tau_2^2\right)\big]\\
 			&\times \exp \big[x_2 \left( \sigma_1 \sigma_2- \tau_1\tau_2 \right)
 			+x_3 \sigma_1+x_4 \sigma_2  +x_5    \big] \\
 			& \times L_k\big[
 			y_1 \left(\sigma_1^2+\tau_1^2\right) + y_2 \left(\sigma_2^2+\tau_2^2\right)\\
 			&+y_3 \left( \sigma_1 \sigma_2- \tau_1\tau_2 \right)
 			+y_4 \sigma_1+y_5 \sigma_2+y_6
 			\big],
 		\end{aligned}
 	\end{equation}
 	where the coefficients $x_i$, $y_i$ and $z_i$ are given as

 	\begin{equation}
 		\begin{aligned}
 			x_1& =-z_1 (\alpha^2 T_S+ \beta^2),  &y_1& = 2z_1 T_S \alpha ^2 ,\\ 
 			x_2& = -4 z_1 \sqrt{T_S} \alpha   \beta,  &y_2&=      4  z_1  \sqrt{T_S}  \alpha \beta \\
 			x_3& =2 i z_1 d \left(\beta   +\alpha  T_S  \right), & y_3&=  2z_1   \beta^2       ,\\
 			x_4& = 2 i z_1 \sqrt{T_S}\, d e^r,  & y_4&= -  2 i z_1 d  \alpha^{-1} \beta e^r  ,\\
 			x_5 &=\frac{z_1}{2}  (T_S-1) d^2 e^{2r}  ,  & y_5&=     2 i z_1 \sqrt{T_S} \,d e^r     ,\\
 			z_1& = [2(\beta^2-\alpha^2 T_S)]^{-1} & y_6&=   \frac{-z_1}{2 \alpha^2} d^2 e^{2r}   ,\\
 			z_2 &= 2 z_1 \alpha^2 (1-T_S) &  &\\ 
 		\end{aligned}
 	\end{equation}
 	with $\alpha =\sinh r$ and  $\beta =\cosh r$.

 	\par
 	\noindent{\bf  Probability of $k$-photon detection\,:}
 	We   evaluate the probability of $k$-photon subtraction   from Eq.~(\ref{reducedcal}) as
 	\begin{equation}\label{app:prob}
 		P_{PS}^{(k)} =\widetilde{ \chi}^{(k)}_{A_1 A'_2}(\Lambda_1, \Lambda_2)  
 		\bigg|_{\substack{\tau_1= \sigma_1=0\\ \tau_2=
 				\sigma_2=0}}=2 z_1 z_2^{k} \exp(x_5) L_k(y_6) .
 	\end{equation}
 	
 	\par
 	\noindent{\bf  Covariance matrix of the  $k$-PSTMSC state\,:}
 	The normalized  Wigner characteristic function   of the  $k$-PSTMSC state  becomes
 	\begin{equation}\label{eq:normalized}
 		\chi^{(k)}_{A_1 A_2'}(\Lambda_1, \Lambda_2)
 		=\left(P_{PS}^{ (k) }\right)^{-1}\widetilde{
 			\chi}^{(k)}_{A_1 A'_2}(\Lambda_1, \Lambda_2) .
 	\end{equation}
 	The average of a symmetrically ordered operator can
 	be calculated by  differentiating the Wigner characteristic
 	function of the $k$-PSTMSC state with respect to $\tau$ and
 	$\sigma$ parameters as follows:
 	\begin{equation}\label{app:covfinalch}
 		\begin{aligned}
 			{}_{\bm{:}}^{\bm{:}}  \hat{q_1}^{r_1} \hat{p_1}^{s_1}
 			\hat{q_2}^{r_2} \hat{p_2}^{s_2} {}_{\bm{:}}^{\bm{:}}   =
 			\widehat{F} \chi^{(k)}_{A_1 A_2'} (\tau_1,
 			\sigma_1,\tau_2, \sigma_2) , 
 		\end{aligned}
 	\end{equation}
 	where 
 	\begin{equation}
 		\begin{aligned}
 			\widehat{F} =&\left( \frac{1}{i} \right)^{r_1+r_2}\left( \frac{1}{-i} \right)^{s_1+s_2}
 			\frac{\partial^{r_1+s_1}}{\partial \sigma_1^{r_1} \partial \tau_1^{s_1} }\\
 			&\times\frac{\partial^{r_2+s_2}}{\partial \sigma_2^{r_2}
 				\partial \tau_2^{s_2} } \{ \bullet  \}_{\substack{\tau_1=
 					\sigma_1=0\\ \tau_2=
 					\sigma_2=0}},
 		\end{aligned}
 	\end{equation}
 	and 
 	${}_{\bm{:}}^{\bm{:}}  \bullet {}_{\bm{:}}^{\bm{:}} $  represents Weyl ordering.
 	We  can choose the values of $r_1$, $s_1$, $r_2$, $s_2$ 
 	in Eq.~(\ref{app:covfinalch}) to yield all the elements of the 
 	covariance matrix. Here we provide the calculated
 	expressions for different moments occurring in the
 	covariance matrix:
 	\begin{equation}\label{firstmoment}
 		\begin{aligned}
 			\langle \hat{q}_1\rangle = -i
 			\left(x_3 -y_4 \frac{L_{k-1}^1\left(y_6\right) }{L_k(y_6)}
 			\right),\\
 		\end{aligned}
 	\end{equation}
 	\begin{equation}
 		\begin{aligned}
 			\langle \hat{q}_2\rangle = -i \left(x_4  -y_5 \frac{L_{k-1}^1\left(y_6\right)  }{L_k(y_6)}  \right),\\
 		\end{aligned}
 	\end{equation}
 	\begin{equation}
 		\begin{aligned}
 			\langle \hat{q}_1^2\rangle = -\left(x_3^2+2 x_1\right)
 			+2 &\left(x_3 y_4+y_1\right)\frac{ L_{k-1}^1\left(y_6\right)}{L_k(y_6)} \\
 			& -y_4^2\frac{ L_{k-2}^2\left(y_6\right) }{L_k(y_6)},\\
 		\end{aligned}
 	\end{equation}
 	\begin{equation}
 		\begin{aligned}
 			\langle \hat{q}_2^2\rangle = -\left(x_4^2+2 x_1\right) +2
 			&\left(x_4 y_5+y_2\right)
 			\frac{L_{k-1}^1\left(y_6\right)}{L_k(y_6)}\\
 			&-y_5^2 \left(  \frac{L_{k-2}^2\left(y_6\right)}{L_k(y_6)}\right),
 		\end{aligned}
 	\end{equation}
 	\begin{equation}
 		\begin{aligned}
 			\langle \hat{p}_1^2\rangle = -2 x_1   +2 y_1 \frac{L_{k-1}^1\left(y_6\right)}{L_k(y_6)},
 		\end{aligned}
 	\end{equation}
 	\begin{equation}
 		\begin{aligned}
 			\langle \hat{p}_2^2\rangle =-2 x_1+ 2 y_2 \frac{L_{k-1}^1\left(y_6\right) }{L_k(y_6)} ,
 		\end{aligned}
 	\end{equation}
 	\begin{equation}
 		\begin{aligned}
 			\langle \hat{q}_1 \hat{q}_2 \rangle =-\left(x_2+x_3 x_4\right) 
 			+&\left(x_4 y_4+x_3 y_5+y_3\right) \frac{L_{k-1}^1\left(y_6\right)}{L_k(y_6)}   \\
 			&-y_4 y_5\frac{ L_{k-2}^2\left(y_6\right)}{L_k(y_6)} ,
 		\end{aligned}
 	\end{equation}
 	\begin{equation}
 		\begin{aligned}
 			\langle \hat{p}_1 \hat{p}_2 \rangle =x_2  -y_3
 			\frac{L_{k-1}^1\left(y_6\right)}{L_k(y_6)} ,
 		\end{aligned}
 	\end{equation}
 	
 	\begin{equation}\label{lastmoment}
 		\begin{aligned}
 			\langle \hat{p}_1 \rangle = \langle \hat{p}_2 \rangle= 
 			\langle \hat{q}_1 \hat{p}_2 \rangle =\langle \hat{p}_1 \hat{q}_2 \rangle=
 			\langle \hat{q}_1 \hat{p}_1 \rangle=\langle \hat{q}_2 \hat{p}_2 \rangle= 0.
 		\end{aligned}
 	\end{equation}
 	Using these moments, the covariance matrix takes the
 	following form:
 	\begin{equation}\label{app:cov}
 		\Sigma_{A_1 A'_2} =
 		\begin{pmatrix}
 			V_{A}^{q} & 0 & V_{C}^{q} & 0 \\
 			0 & V_{A}^{p} & 0 & V_{C}^{p} \\
 			V_{C}^{q} & 0 & V_{B}^{q} & 0 \\
 			0 & V_{C}^{p} & 0 & V_{B}^{p}
 		\end{pmatrix},
 	\end{equation}
 	where $({\Sigma_{A_1 A'_2}})_{ij} =
 	\frac{1}{2}\langle\{ \hat{\xi_i},\hat{\xi_j} \} \rangle -
 	\langle \hat{\xi_i}\rangle \langle \hat{\xi_j}\rangle$. 
 	\subsection{PSTMSC state based CV-MDI-QKD protocol}
 	\label{cvmdipstmsc}
 	\begin{figure*} 
 		\centering
 		\includegraphics[scale=1]{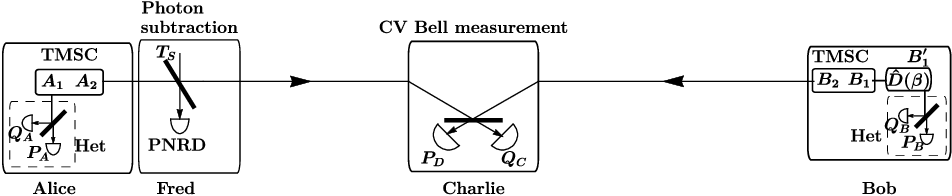}
 		\caption{Schematic of PSTMSC state based
 			CV-MDI-QKD protocol. Alice and Bob start with a TMSC state.
 			Alice sends one of her modes to Fred, an untrusted party,
 			for photon subtraction. The mode is subsequently sent to
 			Charlie, another untrusted party, who then mixes it with the
 			mode received from Bob using a balanced (50-50) beam splitter. He
 			then executes homodyne measurement on the output modes and
 			publicly announces the result. Bob then performs a
 			displacement operation leading to the entanglement of the
 			modes that were retained by Alice and Bob. They then perform
 			heterodyne measurements on their respective modes.  }
 		\label{mdi_scheme}
 	\end{figure*}
 	
 	In the current work, we consider 	PSTMSC state as  resource state for  CV-MDI-QKD.
 The schematic of the CV-MDI-QKD protocol is depicted
 	in Fig.~\ref{mdi_scheme}. The main steps of the protocol are
 	described as follows:

 	\noindent
 	{\bf Step 1:}   Alice starts with a TMSC state in modes $A_1$ and $A_2$ with variance  $V_A = \cosh (2r)$.

 	\noindent
 	{\bf Step 2:} Alice sends one of her modes, $A_2$,
 	to Fred, who performs a photon  subtraction.
 	The photon subtracted mode $A'_2$  is then sent to Charlie
 	through a quantum channel of length $L_{AC}$.
 	
 	\noindent
 	{\bf Step 3:}
 	Bob generates a TMSC state with the two modes denoted by $B_1, B_2$ and   variance $V_B=V_A=V$. He sends the mode $B_2$ to Charlie through another quantum channel of
 	length $L_{BC}$.
 	
 	\noindent
 	{\bf Step 4:}
 	Charlie interferes the two modes 
 	$A'_2$ and $B_2$, obtained from Alice and Bob respectively,
 	using a beam splitter, and the output modes are labeled $C$
 	and $D$.  
 	He then employs homodyne measurements of
 	quadrature $\hat{q}$ on mode $C$  and of  quadrature
 	$\hat{p}$ on mode $D$ to obtain outcomes $\lbrace Q_C,
 	P_D\rbrace$. These outcomes are then publicly announced by
 	Charlie.
 	
 	\noindent
 	{\bf Step 5:} Based on these outcomes, Bob performs a displacement
 	operation $\hat{D}(g(Q_C+ iP_D))$, where $g$  is the gain
 	factor, on his retained mode $B_1$ to obtain the transformed
 	mode $B'_1$.  This operation completes the entanglement
 	swapping process, and the modes $A_1$ and $B'_1$ thus
 	obtained are entangled. Alice and Bob implement heterodyne
 	measurements on their retained modes to obtain outcomes
 	$\lbrace Q_A, P_A\rbrace$ and $\lbrace Q_B, P_B\rbrace$
 	respectively. These outcomes are correlated with each other.

 	\noindent
 	{\bf Step 6:} Finally, Alice and Bob employ classical data
 	post-processing, \ie, information reconciliation (reverse
 	reconciliation) and privacy amplification to distil the
 	secret key.
 	
 	We note that the analysis for the TMSV and PSTMSV states can be obtained as a special case of the PSTMSC state as detailed in next subsection.  
 	
 	\subsection{  Security analysis and equivalent one-way CV-QKD}\label{skr}
 	
 	
 	The CV-MDI-QKD protocol, as shown in
 	Fig.~\ref{mdi_scheme}, uses two quantum channels, one
 	each for
 	transmission of Alice's and Bob's modes to Charlie, and one
 	classical channel, for classical communication. 
 	We assume the two quantum channels to be
 	non-interacting. With this assumption, one-mode entangling
 	cloner collective attacks~\cite{Pirandola-np-2015} on each
 	of the two channels by Eve are possible, while the most
 	general forms of eavesdropping include two-mode correlated
 	attacks~\cite{ottaviani-pra-2015,Pirandola-np-2015}.

\begin{figure}[H]
 		\centering
 		\includegraphics[scale=1]{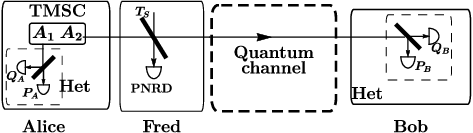}
 		\caption{Schematic of the PSTMSC state based one-way
 			equivalent CV-MDI-QKD protocol.}
 		\label{oneway}
 	\end{figure}

 	In the CV-MDI-QKD protocol described in
 	Fig.~\ref{mdi_scheme}, all of Bob's operations can be
 	assumed to be untrusted, except his heterodyne measurements. This lets us convert the protocol into an equivalent one-way
 	CV-QKD protocol with heterodyne detection (Fig.~\ref{oneway})~\cite{oneway,prl2004,Li-pra-2014}. We note that by
 	calculating the SKR for the equivalent one-way protocol, we
 	obtain a lower bound on the SKR that we would have obtained
 	using the original protocol.
 	Despite this, we use the one-way protocol to calculate the SKR because of convenience in calculations~\cite{winter}. 
 	
 	In the one-way CV-QKD protocol, the total
 	transmission distance from Alice to Bob is $L_{AB}=L_{AC}+L_{BC}$, where $L_{AC}$ and $L_{BC}$ are the transmission
 	distances between Alice to Charlie and Bob to Charlie, which we shall express in kilometer (km).
 	Usually, optical fiber networks are employed as channels for information transfer. The overall loss in the optical fiber between Alice to Charlie for a length $L_{AC}$  is quantified as $l L_{AC}$, where  $l$ is the attenuation factor in decibels per kilometer
 	(dB/km)~\cite{Grasselli2021}.  For this work, we consider $l=0.2$ dB/km~\cite{Liao2017}.  The transmissivity of the corresponding channel is related to its length   via the relation $T_A=10^{-l
 		\frac{L_{AC}}{10}}$. Similarly, the transmissivity for the channel from Bob to Charlie is $T_B=10^{-l
 		\frac{L_{BC}}{10}}$.

 	We will be studying the following two cases in this article:
 	
 	\noindent
 	{\bf Symmetric:} Charlie is considered to be exactly
 	midway between Alice and Bob, so $L_{AC}=L_{BC}$. The total
 	transmission distance $L = 2L_{AC}$ with $T_A = T_B$.
 	
 	\noindent
 	{\bf Extreme asymmetric:}  Charlie is considered to
 	be at the same position as Bob, so $L_{BC} = 0$. The total
 	transmission distance $L = L_{AC}= L_{AB}$ with $T_B = 1$.  
 	
 	The added noise in a quantum channel is defined as
 	\begin{equation}
 		\chi_{channel} = \frac{1-T}{T}+\varepsilon^{\text{th}},
 		\label{eq:chi_line}
 	\end{equation}
 	where $T$ is a normalized parameter for
 	transmissivity and $\varepsilon^{\text{th}}$ is the thermal
 	excess noise from the one-way protocol. $T$ is defined in
 	terms of $T_A$ as
 	\begin{equation}
 		T=\frac{g^2}{2} T_A,
 		\label{eq:T}
 	\end{equation}
 	where $g$ is the gain of the displacement operation
 	employed by Bob. The thermal excess noise is given
 	by~\cite{Ma-pra-2018}
 	\begin{equation}
 		\varepsilon^{th}=\frac{T_B}{T_A}\left(\varepsilon_B^{th} -2\right)+
 		\varepsilon_A^{th}+\frac{2}{T_A},
 		\label{eq:epsilon_thermal}
 	\end{equation}
 	where $\varepsilon_A^{\text{th}}$ and
 	$\varepsilon_B^{\text{th}}$ are the
 	thermal excess noises in the quantum channels from
 	Alice and Bob to Charlie respectively.
 	For $\varepsilon^{\text{th}}$ in Eq.~(\ref{eq:epsilon_thermal}) to be minimum, the gain is
 	\begin{equation}
 		g=\sqrt{\frac{2\left(V_A-1\right)}{T_B\left(V_A+1\right)}}.
 		\label{eq:displacement}
 	\end{equation}
 	
 	The detectors used by Charlie for performing
 	homodyne detection during entanglement swapping are assumed
 	to be imperfect with excess noise given as
 	\begin{equation}
 		\chi_{C}=\frac{v_{el}+1-\eta}{\eta},
 		\label{eq:chi_homo}
 	\end{equation}
 	where, $v_{\text{el}}$ is the electronic noise of the
 	detector with quantum efficiency $\eta$. 
 	The total noise added that is to be considered for the
 	entire setup is then
 	\begin{equation}
 		\chi_{total}=\chi_{channel}+\frac{2 \; \chi_{C}}{T_A}.
 		\label{eq:keyrate}
 	\end{equation}
 	
 	In this article, we have considered both cases of
 	Charlie performing perfect ($\chi_{\text{C}}=0$)  and
 	imperfect homodyne detection. The latter is studied
 	exclusively in Sec.~\ref{subsec:realistic}.
 	
 	The SKR of the protocol described in Sec.~\ref{cvmdi} with
 	channel parameters as given above is
 	\begin{equation}
 		K=P^{(k)}_{PS}\left(\mathcal{R} I_{AB}-\chi_{BE}\right),
 		\label{eq:secure_keyrate}
 	\end{equation}
 	where $I_{AB}$ is the mutual information between
 	Alice and
 	Bob, $\chi_{BE}$ is the Holevo bound between Bob
 	and Eve,   $\mathcal{R}$ is the reconciliation efficiency.
 	We utilize the covariance matrix derived
 	in Eq.~(\ref{app:cov})   for the calculation of SKR
 	for the non-Gaussian PSTMSC state, which provides a lower
 	bound of the SKR according to the optimality of Gaussian
 	attacks~\cite{Cerf-prl-2006}.

 		We start with the calculation of the mutual information.  
 	The covariance matrix of the state $\hat{\rho}_{A_1
 		B'_1}$, obtained after Bob operates the displacement
 	operator $\hat{D}$ on his retained mode $B'_1$ is given by
 	\begin{equation}
 		\begin{aligned}
 			\Sigma_{A_1 B'_1} = &\begin{pmatrix}
 				V_{A}^{q} & 0 & \sqrt{T}V_{C}^{q} & 0 \\
 				0 & V_{A}^{p} & 0 & \sqrt{T}V_{C}^{p} \\
 				\sqrt{T}V_{C}^{q} & 0 & T(V^q_{B}+\chi_{total}) & 0 \\
 				0 & \sqrt{T}V_{C}^{p} & 0 & T(V^p_{B}+\chi_{total})
 			\end{pmatrix},\\
 			&=\begin{pmatrix}
 				\delta_1 & 0 & \kappa_1 & 0 \\
 				0 &\delta_2 & 0 & \kappa_2 \\
 				\kappa_1 & 0 & \mu_1 & 0 \\
 				0 & \kappa_2& 0 & \mu_2 \\
 			\end{pmatrix}
 			=\begin{pmatrix}
 				\Sigma_{A_1} & \Sigma_{C} \\
 				\Sigma_{C} & \Sigma_{B'_1}\\
 			\end{pmatrix}.
 			\label{eq:variance_a1b1}
 		\end{aligned}
 	\end{equation}
 	
 	The mutual information between Alice and Bob is given
 	as~\cite{Laudenbach-review}
 	\begin{equation}
 		I_{AB}=\frac{1}{2}\log_2\left(\frac{\Sigma_{A_1}^q+1}
 		{\Sigma_{{A_1}|B'_1}^q+1}\right)+
 		\frac{1}{2}\log_2\left(\frac{\Sigma_{A_1}^p+1}
 		{\Sigma_{{A_1}|B'_1}^p+1}\right),
 		\label{eq:mutualinfo}
 	\end{equation}
 	where $\Sigma_{{A_1}|B'_1}$ is the conditional
 	variance of the outcome of Alice depending on Bob's
 	heterodyne measurement results and can be evaluated from 
 	\begin{equation}
 		\begin{aligned}
 			\Sigma_{A_1|B'_1} =
 			&\Sigma_{A_1}-\Sigma_C\left(\Sigma_{B'_1}+
 			\mathbb{1}_2\right)^{-1}(\Sigma_C)^T,\\
 			=&\begin{pmatrix}
 				\delta_1-\frac{\kappa_1^2}{\mu_1+1} & 0 \\
 				0 & \delta_2-\frac{\kappa_2^2}{\mu_2+1} \\
 			\end{pmatrix}.
 		\end{aligned}
 		\label{eq:varianceagivenb}
 	\end{equation}

 	To calculate the Holevo bound, we assume that Eve holds a
 	purification of the state $\hat{\rho}_{A_1B'_1EF}$ having
 	access to Fred's mode.
 	The Holevo bound between Bob and Eve is
 	\begin{equation}
 		\begin{aligned}
 			\chi_{BE} = S(\hat{\rho}_{A_1 B'_1}) - S(\hat{\rho}_{A_1|B'_1}),
 			\label{eq:holevo}
 		\end{aligned}
 	\end{equation}
 	where $S(\hat{\rho})$ is the von-Neumann entropy for the
 	state $\hat{\rho}$ and is given by
 	\begin{equation}
 		S(\hat{\rho}) = \Sigma_i g(\lambda_i),
 		\label{eq:rho}
 	\end{equation}
 	where,
 	\begin{equation}
 		g(\lambda) =
 		\frac{\lambda+1}{2}\log_2\left(\frac{\lambda+1}{2}\right)-
 		\frac{\lambda-1}{2}\log_2\left(\frac{\lambda-1}{2}\right),
 		\label{eq:rho_g}
 	\end{equation}
 	and $\lambda_i$ are the symplectic eigenvalues of the corresponding matrices.
 	$S(\hat{\rho}_{A_1 B'_1})$ is calculated using the
 	symplectic eigenvalues of the matrix in
 	Eq.~(\ref{eq:variance_a1b1}) which evaluate as
 	\begin{equation}
 		\begin{aligned}
 			\lambda_{1,2} = \frac{1}{\sqrt{2} }\left[X \pm \sqrt{X^2-4 Y}\right]^{1/2},
 		\end{aligned}
 	\end{equation}
 	with $X = \delta_1 \delta_2+\mu_1 \mu_2+2\kappa_1 \kappa_2$ and $Y = (\delta_1
 	\mu_1-\kappa_1^2)(\delta_2 \mu_2-\kappa_2^2)$, while
 	$S(\hat{\rho}_{A_1|B'_1})$ is calculated using the
 	symplectic eigenvalue of the matrix in
 	Eq.~(\ref{eq:varianceagivenb}) which evaluates as
 	\begin{equation}
 		\lambda_3 = \sqrt{\left(
 			\delta_1-\frac{\kappa_1^2}{\mu_1+1}\right)\left(
 			\delta_2-\frac{\kappa_2^2}{\mu_2+1}\right)}.
 	\end{equation}
 	Using these expressions, secret key rate~(\ref{eq:secure_keyrate}) can be readily evaluated.

 	\noindent
 	{\bf Special cases:} On setting $d=0$, we obtain the SKR for the PSTMSV state. Further, in the unit transmissivity limit $T_S \rightarrow1$ with $d=0$ and $k=0$, we obtain the SKR for TMSV state.  
 	
 	\section{Optimization of state parameters}\label{sec:opt}
 	After describing the CV-MDI-QKD protocol in detail, we now
 	proceed to optimize the state parameters in order to
 	maximize the SKR.  We note that, for PSTMSC states, the
 	state parameters are variance, displacement, and
 	transmissivity.  Similarly, for the PSTMSV state, which is
 	obtained when displacement is zero, the state parameters
 	reduce to only variance and transmissivity. We consider two
 	different optimization scenarios for the SKR: (i)
 	optimization of SKR with respect to displacement $d$ and
 	transmissivity $T_S$  at a fixed variance, (ii) optimization
 	of SKR with respect to all the state parameters $V$, $d$,
 	and $T_S$.

 	\subsection{Optimization of  $d$ and $T_S$}
 	We consider the  optimization of  displacement and
 	transmissivity, at a fixed variance, in order to maximize
 	the SKR for the PSTMSC resource state.  The displacement
 	range has been set to $d \in (0,5)$, as increasing beyond it
 	only improves the key rate slightly, but the average number
 	of photons increases significantly which requires a
 	high-powered source for preparing the resource state. For
 	the PSTMSV resource state, we need to only optimize with
 	respect to transmissivity, at a fixed variance, to maximize
 	the SKR.   TMSV state with a maximum squeezing of 15 dB has been
 	achieved~\cite{vahlbruch-prl-2016}, which is equivalent  to
 	a squeezing of $r=1.73$ or a variance of $V=\cosh \, 2r =
 	15.83$.  We, therefore,
 	optimize the above parameters at   three fixed variance
 	values $(V = 5$, $10$, and $15)$ to maximize the SKR: 
 	\begin{equation}
 		\begin{aligned}
 			\max_{d, T_S} \quad & K(V,d,T_S)\\
 			\textrm{s.t.} \quad & 0\leq d \leq 5,\\
 			&0\leq T_S \leq 1.  \\
 		\end{aligned}
 	\end{equation}
 	We note that the optimization of state parameters is
 	performed for each value of transmission distance.

 	\begin{figure}[htbp] 
 		\centering
 		\includegraphics[scale=1]{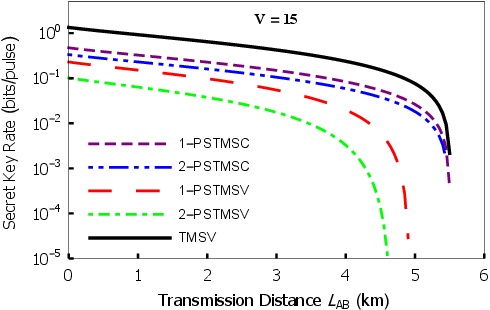}
 		\caption{[Color online]  Secret key rate
 			(SKR) 
 			as a function of  transmission
 			distance for fixed variance  $(V = 15)$ in the symmetric
 			case  ($L_{AB}=2L_{AC}=2L_{BC} $ )  using perfect homodyne
 			detectors $(\eta = 1, \nu_{\text{el}} = 0)$. The SKR has
 			been optimized with respect to displacement  and
 			transmissivity. Here, reconciliation efficiency
 			$\mathcal{R}=0.96$,  and excess noise
 			$\epsilon_A^{\text{th}}=\epsilon_B^{\text{th}}=0.002$. }
 		\label{keysym2}
 	\end{figure}
 	
 	We first consider the symmetric case, \ie, Charlie
 	is based between Alice and Bob. It has been shown in earlier
 	works that the symmetric case yields an extremely low value
 	of maximum transmission distance~\cite{Li-pra-2014,
 		Pirandola-np-2015}; therefore, we analyze this case briefly
 	for only $V = 15$. 
 	
 	We plot the SKR as a function of  transmission
 	distance  for $V = 15$ in Fig.~\ref{keysym2}. As we have
 	mentioned earlier, for the symmetric case,  the maximum
 	transmission distance of TMSV falls around $5.5$ km. The
 	TMSV, 1-PSTMSC, and 2-PSTMSC states yield the same maximum
 	transmission distance. However, the SKR for the 1-PSTMSC and
 	2-PSTMSC states is less as compared to the TMSV state SKR.
 	We also observe that 1-PSTMSV and 2-PSTMSV states yield
 	smaller maximum transmission distance and SKR as compared to
 	all the aforementioned states.
 	
 	\begin{figure}[htbp] 
 		\centering
 		\includegraphics[scale=1]{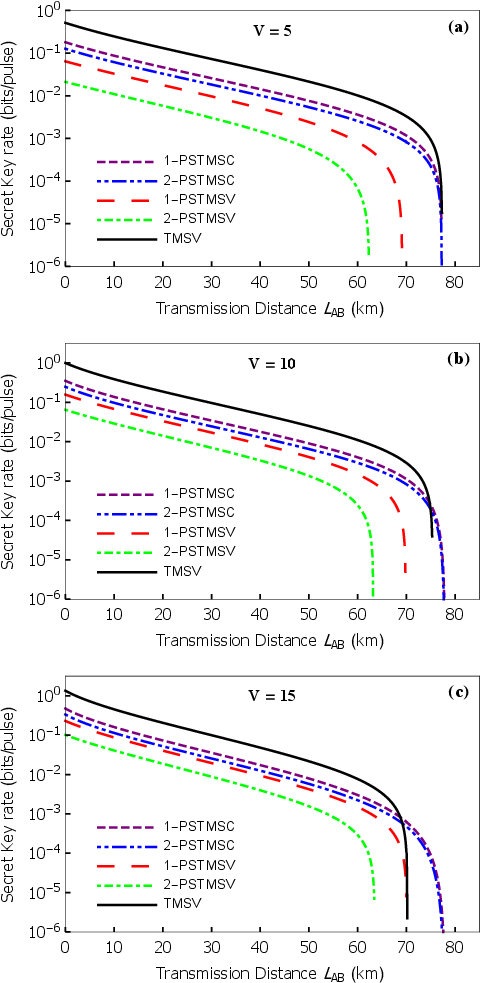}
 		\caption{[Color online]  Secret key rate
 			(SKR) as a function of  transmission distance for fixed
 			variances:  (a) $V = 5$, (b) $V = 10$, (c) $V = 15$, in the
 			extreme asymmetric case  ($L_{BC} = 0$ km)   using perfect
 			homodyne detectors $(\eta = 1, \nu_{\text{el}} = 0)$. The
 			SKR has been optimized with respect to displacement  and
 			transmissivity.  Here, reconciliation efficiency
 			$\mathcal{R}=0.96$, and excess noise
 			$\epsilon_A^{\text{th}}=\epsilon_B^{\text{th}}=0.002$. }
 		\label{keypartialfinal}
 	\end{figure}
 	
 	We now proceed to analyze the extreme asymmetric
 	case, where Charlie is located close to Bob, \ie, $L_{BC} =
 	0$ km. Figure~\ref{keypartialfinal} shows the plot of SKR as
 	a function of  transmission distance for three different
 	fixed variances:  $V = 5$,  $ 10$, and  $15$.
 	The results show that while the photon-subtracted
 	states yield a constant maximum transmission distance with
 	changing variance; for the TMSV state, the maximum
 	transmission distance decreases with increasing variance.
 	This indicates a positive role played by photon subtraction.
 	We now stress on the role of displacement through two
 	specific points: (i) The maximum transmission distance and
 	the SKR achieved by the 1-PSTMSC and 2-PSTMSC states
 	outperform the respective values of both the PSTMSV states
 	(ii) In addition, we notice that, unlike the PSTMSV states,
 	the maximum transmission distance achieved by 1-PSTMSC and
 	2-PSTMSC states is almost same, which suggests that
 	displacement might be capable of maintaining the same
 	maximum transmission distance for multiphoton subtracted
 	states. This interesting fact is explored in the next
 	section and exploited for a novel protocol that ensures
 	maximum usage of resources.

 	For variance $V=5$, the maximum transmission
 	distance of 1-PSTMSC and 2-PSTMSC states is almost the same
 	as the TMSV state; however, the secret key rate for the TMSV
 	state is higher as compared to 1-PSTMSC and 2-PSTMSC states.
 	Both 1-PSTMSV and 2-PSTMSV states underperform in terms of
 	SKR and maximum transmission distance as compared to the
 	aforementioned states. For variance $V=10$  and $V=15$, the maximum
 	transmission distance of 1-PSTMSC and 2-PSTMSC states is
 	higher compared to the TMSV state, but with a slight decrease in the SKR.
 	On comparison with the results of earlier works, the optimization of parameters not only leads to the enhancement of
 	maximum transmission distance, but also a substantial
 	improvement in SKR.

 	\begin{figure}[htbp] 
 		\centering
 		\includegraphics[scale=1]{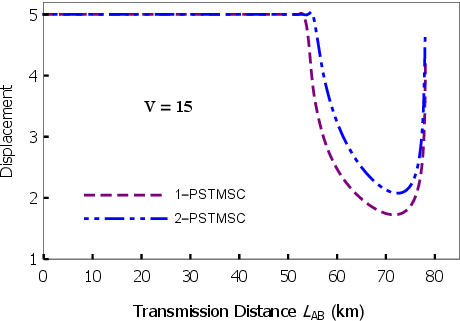}
 		\caption{[Color online]  Optimal
 			displacement for the maximal secret key rate (SKR) as a
 			function of  transmission distance for a fixed variance  $(V
 			= 15)$ in the extreme asymmetric case using   perfect
 			homodyne detectors $(\eta = 1, \nu_{\text{el}} = 0)$. Here,
 			reconciliation efficiency  $\mathcal{R}=0.96$,  and excess noise
 			$\epsilon_A^{\text{th}}=\epsilon_B^{\text{th}}=0.002$. }
 		\label{dpartialfinal}
 	\end{figure}
 	
 	In Fig~\ref{dpartialfinal}, we show the optimal 
 	displacement as a function of transmission distance,
 	which maximizes the SKR for 1-PSTMSC and 2-PSTMSC states for
 	$V = 15$ (Fig~\ref{keypartialfinal}(c)).  The displacement
 	remains equal to $5$ for transmission distances below
 	$L_{AB} \approx 50$ km, after which it starts to decrease.
 	As the transmission distance is increased, it attains a
 	minima and then has a steep rise. We notice that the
 	displacement required to obtain maximal SKR for a 1-PSTMSC
 	state is lower than the 2-PSTMSC state.

 	\begin{figure}[htbp] 
 		\centering
 		\includegraphics[scale=1]{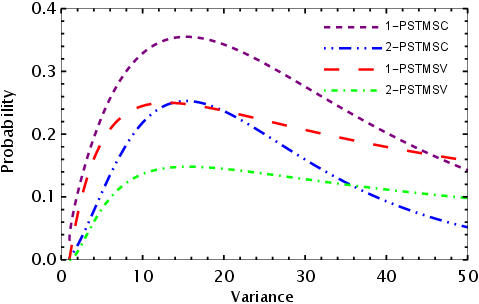}
 		\caption{[Color online] Probability  of successful photon
 			subtraction as a function of variance.
 			Optimal displacement  and transmissivity for the extreme
 			asymmetric case using perfect homodyne detectors $(\eta = 1,
 			\nu_{\text{el}} = 0)$ at transmission distance of $L_{AB} =
 			60$ km have been considered. }
 		\label{probfinal}
 	\end{figure}  
 	We plot the probability of successful photon
 	subtraction as a function of variance in
 	Fig.~\ref{probfinal}, where the optimal displacement  and
 	transmissivity which maximize the SKR for $V = 15$
 	(Fig~\ref{keypartialfinal}(c)) and a transmission distance
 	of $L_{AB} = 60$ km   have been taken. 
 	The magnitude of probability can also be interpreted
 	as a measure of resource usage. 
 	At $V=15$, the probabilities for 1-PSTMSC, 2-PSTMSC, 1-PSTMSV and 
 	2-PSTMSV states are $0.33$, $0.22$, $0.23$,
 	and $0.15$, respectively. Therefore, maximum resource usage
 	occurs for the  1-PSTMSC state.

 	\subsection{Post selection of photon subtracted state}
 	
 	As explicit from Fig.~\ref{keypartialfinal}, the
 	maximum transmission distance of 2-PSTMSC is the same as
 	1-PSTMSC. We  explicitly provide the maximum transmission
 	distances of up to 4-PSTMSC and 4-PSTMSV states in
 	table~\ref{table1}. We observe that all the $k$-PSTMSC
 	states have almost the same maximum transmission distances,
 	while for PSTMSV states, the maximum transmission distance
 	decreases as $k$ is increased.
 	
 	\begin{table}[H]
 		\caption{\label{table1} Maximum transmission
 			distance $d_{\text{max}}$   for $k$-PSTMSC and $k$-PSTMSV
 			states.}
 		\renewcommand{\arraystretch}{1.5}
 		\begin{center}
 			\begin{tabular}{ |c|c|c|c|} 
 				\hline \hline
 				State & $d_{\text{max}}$ (in km)  & State & $d_{\text{max}}$ (in km) \\ \hline
 				1-PSTMSV & 69.9  & 1-PSTMSC & 76.8 \\ \hline 
 				2-PSTMSV & 63.3  & 2-PSTMSC & 76.5 \\ \hline 
 				3-PSTMSV & 57.3  & 3-PSTMSC & 76.3 \\ \hline 
 				4-PSTMSV & 51.8  & 4-PSTMSC & 76.1\\ 
 				\hline \hline
 			\end{tabular}
 		\end{center}
 	\end{table}
 	
 	Therefore, we can envisage a new QKD protocol, where
 	we can use different photon subtracted TMSC states for key
 	generation.
 	To this end, Fred uses a PNRD, which can detect up
 	to $N$ photons. The positive operator valued measure (POVM)
 	of such a PNRD can be written as 
 	\begin{equation}
 		\bigg\{\Pi_0,\Pi_1,\dots \Pi_k, 
 		\dots, \Pi_N,\mathbb{1}-\sum_{k=0}^N \Pi_k\bigg\}, \quad \Pi_k = |k\rangle \langle k|.
 	\end{equation}
 	On detection of $k$ photons, Fred announces publicly
 	that $k$ photons have been subtracted. This information is
 	used to construct an effective information channel for each
 	$k$~\cite{Silberhorn-prl-2002}. Further steps in the QKD
 	protocol including classical post-processing on each of
 	these effective information channel is done
 	separately~\cite{Dequal-npj-2021}. Denoting the SKR for  the
 	$k$-PSTMSC state by $K(\hat{\rho}^{(k)})$, the average SKR
 	over different  information channels can be written
 	as~\cite{scarani-rmp-2009}
 	\begin{equation}
 		K_{\text{avg}} = \sum_{k=1}^N K(\hat{\rho}^{(k)}).
 	\end{equation}
 	We note that we have excluded zero-photon catalyzed
 	TMSC state ($k=0$) from the calculation of $K_{\text{avg}}$.
 	Further, we have restricted the range of $k$ to be $k \in
 	(1, 4)$, \ie, $N_\text{max}=4$.
 	\begin{figure}[htbp] 
 		\centering
 		\includegraphics[scale=1]{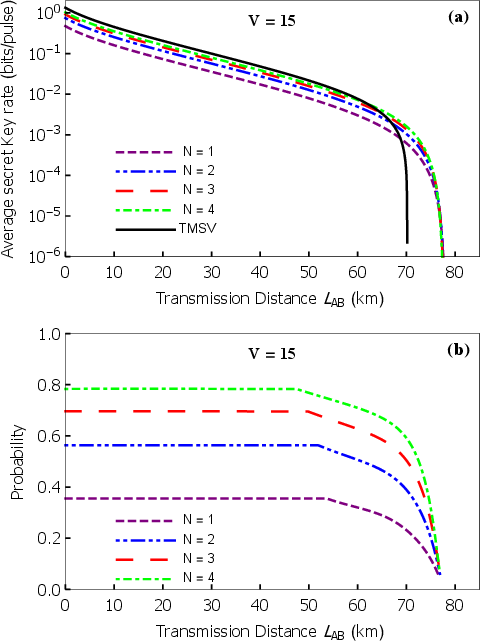}
 		\caption{[Color online] (a) Average secret
 			key rate (SKR) as a function of  transmission distance for
 			fixed variance $V = 15$, in the extreme asymmetric case
 			($L_{BC} = 0$ km)  using perfect homodyne detectors $(\eta =
 			1, \nu_{\text{el}} = 0)$. $N$ represents the contribution in
 			the SKR obtained from 1-PSTMSC, 2-PSTMSC, $\dots$,
 			$N$-PSTMSC states. The SKR has been optimized with respect
 			to displacement  and transmissivity.  Here, reconciliation
 			efficiency $\mathcal{R}=0.96$, and excess noise
 			$\epsilon_A^{\text{th}}=\epsilon_B^{\text{th}}=0.002$. 
 			(b) Probability as a function of
 			transmission distance. Probability has been calculated for
 			optimum  displacement  and transmissivity at every
 			transmission distance. }
 		\label{post}
 	\end{figure}

 	Figure~\ref{post}(a) shows the plot of the SKR as a
 	function of transmission distance for $N=1,2,3$, and $4$. We
 	observe that while there is no enhancement in maximum
 	transmission distance, the SKR does show an improvement.
 	Further, as the value of $N$ is increased, the SKR also
 	increases. We plot the total probability $P_{\text{total}} =
 	\sum_{k=1}^N P(\hat{\rho}^{(k)})$ as a function of the
 	transmission distance in Fig.~\ref{post}(b) for $N=1,2,3$,
 	and $4$. The total probability reaches up to a maximum of $
 	0.36$, $0.56$, $ 0.69$, and $0.78$  for $N=1,2,3$, and $4$,
 	respectively. Thus,  we see that as   $N$ is increased, the
 	total probability increases. As mentioned earlier, the
 	probability of photon subtraction is a measure of the
 	resource state utilization per trial. The enhancement of the
 	total probability indicates an efficient utilization of the
 	resource state providing an  enhancement in the SKR in our
 	new scheme.

 	\subsection{Optimization of $V$,  $d$ and $T_S$ }

\begin{figure}[htbp] 
 		\centering
 		\includegraphics[scale=1]{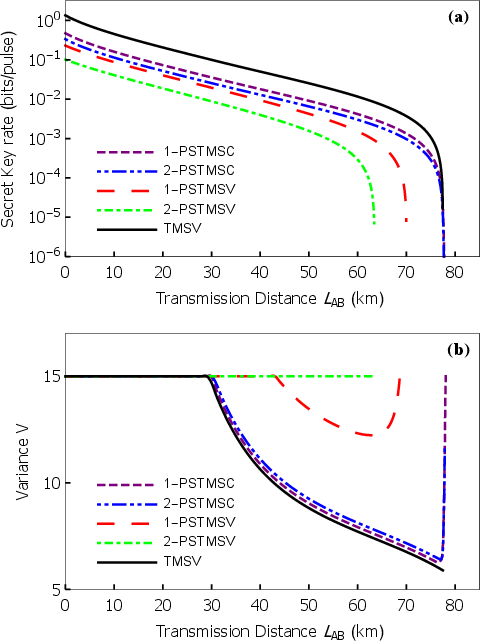}
 		\caption{[Color online]  (a) Secret key rate
 			(SKR) as a function of  transmission distance in the extreme
 			asymmetric case using perfect homodyne detectors $(\eta = 1,
 			\nu_{\text{el}} = 0)$. The SKR has been optimized with
 			respect to variance, displacement, and transmissivity. (b)
 			Optimal variance maximizing the SKR  as a function of
 			transmission distance. Here, reconciliation efficiency
 			$\mathcal{R}=0.96$,  and excess noise
 			$\epsilon_A^{\text{th}}=\epsilon_B^{\text{th}}=0.002$.}
 		\label{allfinal}
 	\end{figure}

 	In this subsection, we optimize the variance,
 	displacement, and transmissivity for various states in order
 	to maximize the SKR:
 	\begin{equation}
 		\begin{aligned}
 			\max_{V,d, T_S} \quad & K(V,d,T_S)\\
 			\textrm{s.t.} \quad & 1\leq V \leq 15,\\
 			& 0\leq d \leq 5,\\
 			&0\leq T_S \leq 1.  \\
 		\end{aligned}
 	\end{equation} We plot the SKR as a function of
 	transmission distance in the extreme asymmetric case using
 	perfect homodyne detectors in Fig.~\ref{allfinal}(a).  
 	We observe that the maximum transmission distance of
 	TMSV and PSTMSC states are the same; however, the SKR for
 	the TMSV state is higher as compared to  PSTMSC states. The
 	PSTMSV states  underperform in terms of both the SKR and the
 	maximum transmission distance as compared to the
 	aforementioned states.  Therefore, photon subtraction
 	operations do not enhance the SKR or the maximum
 	transmission distance.

 	The optimized variance as a function of transmission
 	distance is shown in Fig.~\ref{allfinal}(b). We note that
 	the SKR is maximized at the maximum allowed variance $V=15$
 	up to a transmission distance $L_{AB} \approx 30$ km for
 	TMSV and PSTMSC states. The optimal variance tends to
 	decrease as the transmission distance is increased and
 	reaches a minimum value of $V\approx 6$ at the maximum
 	transmission distance.
 	For the 1-PSTMSV states, the optimal variance
 	remains equal to $15$ for larger transmission distances. The
 	optimal variance decreases, attains minima, and then starts
 	increasing as the transmission distance is increased
 	further. For the 2-PSTMSV states, the optimal variance
 	remains at $V=15$ for all transmission distances.
 	
 	To conclude this section, we find that neither
 	PSTMSC nor PSTMSV state provides any advantage over the TMSV
 	state when the optimization is performed with respect to all
 	state parameters. We obtain enhanced performance at fixed
 	high variance but as states with small variance are easy to
 	prepare, it is better to work with TMSV or TMSC state.

 	\subsection{Imperfect detectors}
 	\label{subsec:realistic}

\begin{figure}
 		\centering
 		\includegraphics[scale=1]{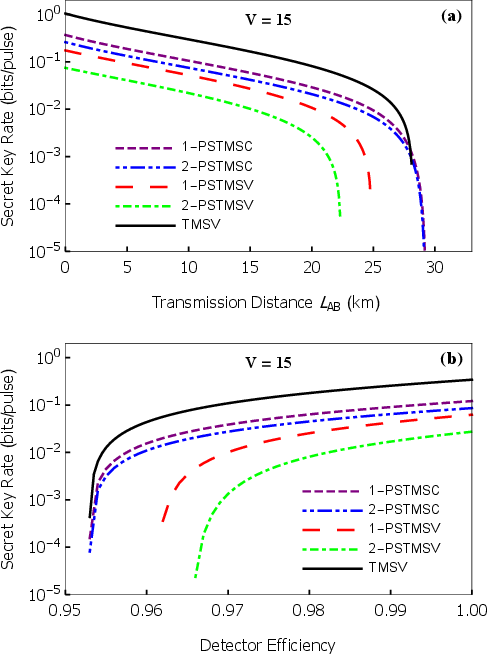}
 		\caption{[Color online] (a) Secret key rate
 			(SKR) as a function of  transmission distance for fixed
 			variance $(V=15)$ in the extreme asymmetric case using
 			imperfect homodyne detectors  $(\eta = 0.995,
 			\nu_{\text{el}} = 0.01)$. (b) Secret key rate (SKR) as a
 			function of the efficiency of the homodyne detector. The SKR
 			has been optimized with respect to displacement and
 			transmissivity. Here, reconciliation efficiency
 			$\mathcal{R}=0.96$,  and excess noise
 			$\epsilon_A^{\text{th}}=\epsilon_B^{\text{th}}=0.002$. The
 			transmission distance has been fixed at $10$ km.}
 		\label{eta1v}
 	\end{figure}

 	We now consider the case of imperfect homodyne
 	detectors at fixed variance $V = 15$. In
 	Fig.~\ref{eta1v}(a), we plot the SKR, optimized with respect
 	to  displacement and transmissivity, as a function of
 	transmission distance in the extreme asymmetric case. The
 	efficiency and the electronic noise corresponding to
 	Charlie's homodyne detectors are set as $\eta = 0.995,
 	\nu_{\text{el}} = 0.01$.
 	We observe that the transmission distance reduces
 	significantly from $L_{AB} = 80$ km
 	(Fig.~\ref{keypartialfinal}(c)) to $L_{AB} = 30$ km.
 	Further, the difference in the maximum transmission
 	distance between TMSV and 1-PSTMSC state reduces from 8 km
 	(Fig.~\ref{keypartialfinal}(c)) to 1 km. 
 	Furthermore, the maximum transmission distance for
 	the PSTMSV states is lower than the TMSV state.

 	We plot the SKR as a function of the efficiency of
 	the homodyne detector in Fig.~\ref{eta1v}(b). We see that
 	the TMSV state is most robust to detector noise, followed by
 	the PSTMSC states. To provide numerical values, the TMSV and
 	PSTMSC states can yield SKR up to a threshold detector
 	efficiency of $\eta = 0.953$. Further, for the 1-PSTMSV and
 	2-PSTMSV states, the threshold efficiency increases to $\eta
 	= 0.962$ and $\eta = 0.966$, respectively.

 	\begin{figure}
 		\centering
 		\includegraphics[scale=1]{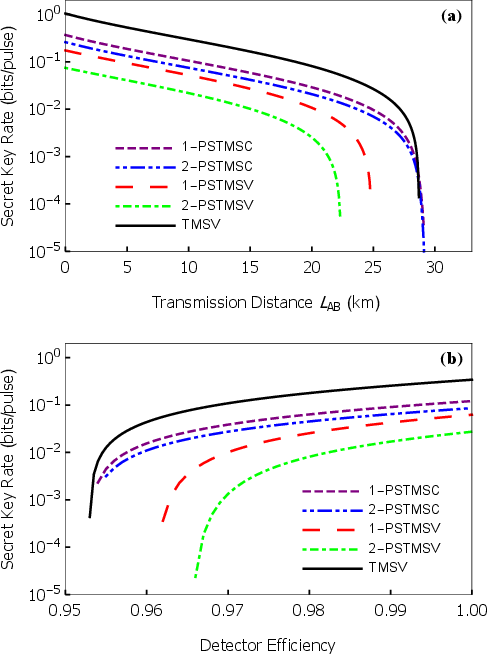}
 		\caption{[Color online] (a) Secret key rate
 			(SKR) as a function of  transmission distance  in the
 			extreme asymmetric case using imperfect homodyne detectors
 			$(\eta = 0.995, \nu_{\text{el}} = 0.01)$. (b) SKR as a
 			function of  the efficiency of the homodyne detector. The
 			SKR  has been optimized with respect to variance,
 			displacement, and transmissivity. Here, reconciliation
 			efficiency  $\mathcal{R}=0.96$,  and excess noise
 			$\epsilon_A^{\text{th}}=\epsilon_B^{\text{th}}=0.002$. The
 			transmission distance has been fixed at $10$ km.}
 		\label{eta1all}
 	\end{figure}

 	In Fig.~\ref{eta1all}(a), we plot the SKR optimized
 	with respect to variance, displacement, and transmissivity
 	as a function of  transmission distance for  the extreme
 	asymmetric case. We observe that the PSTMSC state
 	outperforms the TMSV state in terms of maximum transmission
 	distance by a very small margin but underperforms in terms
 	of the SKR. On the other hand, the PSTMSV states have
 	smaller transmission distances than the TMSV state.
 	
 	The plot of the SKR as a function of the efficiency
 	of the homodyne detector is shown in Fig.~\ref{eta1all}(b).
 	It can be seen that the  TMSV and PSTMSC states can yield an
 	SKR up to a threshold detector efficiency of $\eta \approx
 	0.954$. Further, for the 1-PSTMSV and 2-PSTMSV states, the
 	threshold efficiency increases to $\eta = 0.962$ and $\eta =
 	0.966$, respectively.

 	\section{Conclusion}
 	\label{sec:conc}
 	In this work, we critically examined the benefits of
 	the PSTMSV and PSTMSC resource states in CV-MDI-QKD as
 	claimed in Refs.~\cite{Ma-pra-2018,chandan-pra-2019}. To
 	this end, we derive the Wigner characteristic function of
 	the PSTMSC resource states, which is then used to derive its
 	covariance matrix. 
 	In the first scenario, the evaluated SKR is
 	maximized with respect to displacement and transmissivity
 	for various fixed variances. This significantly improved the
 	performance of PSTMSC states and  established its
 	superiority over the PSTMSV states in terms of both SKR and
 	maximum transmission distance up to which secure QKD can be
 	implemented.    Although PSTMSV states are never superior when compared to TMSV states, PSTMSC states only offer an advantage over TMSV states under conditions of high variance.

 	In the second scenario, where we perform
 	optimization of the SKR with respect to  variance,
 	displacement, and transmissivity, we find that the photon
 	subtraction operation on Alice's side does not provide any
 	benefits over the TMSV state. The optimal variance is
 	relatively small at large transmission distances. Since
 	preparing states with low variances is less challenging, our
 	work conclusively establishes that performing PS operation
 	by Alice on TMSV and TMSC states is unessential, unlike the
 	conclusion of earlier works for PSTMSV and PSTMSC state.
 	
   In this work, we have focused on photon subtraction only on Alice's side. It is essential to consider photon subtraction on Bob's side, as well as its 
 		advantage in other CV-QKD protocols~\cite{middle,ebcvqkd,virtual16} needs to be assessed. Our work shows that photon subtraction on Alice's side for CV-MDI-QKD  is unessential and  raises numerous questions regarding the utility of non-Gaussian operations 
 		in CV quantum information processing, which is currently pursued by a large community of researchers. 
 		
 		Photon subtraction  has been shown to be useful in quantum
 		teleportation~\cite{tel2000,dellanno-2007,tel2009,catalysis15,catalysis17,wang2015,tele-2023,noisytele}
 		and quantum metrology~\cite{gerryc-pra-2012,josab-2012,braun-pra-2014,josab-2016,pra-catalysis-2021,ill2008,ill2013,metro22,metro-thermal-arxiv,ngsvs-arxiv}. The potential benefits of photon subtraction and other non-Gaussian 
 		operations such as photon catalysis~\cite{catpra2019,virtualzpccat} and photon addition~\cite{addition}
 		in CV-QKD and other CV-QIP tasks demand a thorough examination.

 	We proposed a new protocol that maximizes resource
 	utilization by exploiting the fact that multiphoton
 	subtracted  TMSC state yields the same maximum transmission
 	distance as that of 1-PSTMSC. This protocol does not provide
 	any improvements in the maximum transmission distance, but
 	the SKR   increases as compared to 1-PSTMSC resource state. 
 	It should be stressed that this protocol is effective at high variance. 
 	
 	We have also derived the Wigner characteristic function of the PSTMSC
 	states, which can be helpful in studying
 	nonclassicality~\cite{Hillery,Mandel:79,antibunching},
 	nonlocality~\cite{am2002,nonlocality}, and
 	non-Gaussianity~\cite{non-G} as well as can be employed in
 	various other QIP protocols such as entanglement swapping,
 	quantum teleportation, and quantum metrology.

 	\section*{Acknowledgement}
 	S.C. acknowledges the Prime Minister's Research Fellowship (PMRF) scheme, GoI, for financial support. A and C.K. acknowledge  the financial
 	support from {\bf DST/ICPS/QuST/Theme-1/2019/General} Project
 	number {\sf Q-68}.

 	\appendix
 	\section{Phase space formalism of  CV  systems and Wigner
 		characteristic function for the PSTMSC state}
 	\label{app:char}
 	In this section, we give a brief overview of CV systems and
 	their description using Wigner characteristic functions. We
 	then provide a detailed calculation of the Wigner
 	characteristic function for PSTMSC states, the corresponding
 	success probability, and the covariance matrix.
 	\subsection{CV systems}
 	\label{cvsystem}
 	An $n$-mode continuous variable 
 	quantum system can be represented via $n$
 	pairs of Hermitian quadrature
 	operators~\cite{arvind1995,Braunstein,adesso-2007,Weedbrook-rmp-2012,adesso-2014}
 	\begin{equation}\label{eq:columreal}
 		\hat{ \xi}  =(\hat{ \xi}_i)= (\hat{q_{1}},\,
 		\hat{p_{1}} \dots, \hat{q_{n}}, 
 		\, \hat{p_{n}})^{T}, \quad i = 1,2, \dots ,2n.
 	\end{equation}
 	The   commutation relation  between the quadrature operators
 	can be succinctly expressed in shot noise units ($\hbar$=2)
 	as 
 	\begin{equation}\label{eq:ccr}
 		[\hat{\xi}_i, \hat{\xi}_j] = 2 i \Omega_{ij}, \quad (i,j=1,2,...,2n),
 	\end{equation}
 	where $\Omega$ is given by 
 	\begin{equation}
 		\Omega = \bigoplus_{k=1}^{n}\omega =  \begin{pmatrix}
 			\omega & & \\
 			& \ddots& \\
 			& & \omega
 		\end{pmatrix}, \quad \omega = \begin{pmatrix}
 			0& 1\\
 			-1&0 
 		\end{pmatrix}.
 	\end{equation}

 	Alternatively, an $n$-mode continuous variable 
 	quantum system can also be represented via $n$-pairs
 	of annihilation  $\hat{a}_i$ and creation operators
 	$\hat{a}^{\dagger}_i$ $(i=1,2,...,n)$.
 	The operators $\hat{a}_i\, \text{and}\, {\hat{a}_i}
 	^{\dagger}$are related via the quadrature operators as 
 	\begin{equation}\label{realtocom}
 		\hat{a}_i=   \frac{1}{2}(\hat{q}_i+i\hat{p}_i),
 		\quad  \hat{a}^{\dagger}_i= \frac{1}{2}(\hat{q}_i-i\hat{p}_i).
 	\end{equation}

 	The displacement operator displacing the $\hat{q}_i$ and
 	$\hat{p}_i$ quadratures in phase space by an amount
 	$q_i$ and $p_i$, respectively, is defined  as 
 	\begin{equation}\label{dis}
 		\hat{D}_i(q_i,p_i) = e^{i(p_i\hat{q}_i-q_i
 			\hat{p}_i)}.
 	\end{equation}
 	
 	Symplectic transformations are the linear homogeneous transformations characterized by real
 	$2n \times 2n$ matrices $S$, and act on the quadrature
 	operators as $\hat{\xi}_i \rightarrow
 	\hat{\xi}_i^{\prime} = S_{ij}\hat{\xi}_{j}$. 
 	Symplectic matrices satisfy the condition $S\Omega
 	S^T = \Omega$, which can be obtained from the commutation
 	relations~(\ref{eq:ccr}).
 	We define below two important symplectic transformation
 	which represent the  beam splitter action and two-mode squeezer,
 	which are relevant to this work.
 	\par
 	\noindent{\bf Beam splitter\,:}
 	The action of a beam splitter on the quadrature operators 
 	$  \hat{\xi} = (\hat{q}_{i}, \,\hat{p}_{i},\, \hat{q}_{j},\,
 	\hat{p}_{j})^{T}$ of a two-mode system 
 	is defined through the following  transformation matrix:
 	\begin{equation}\label{beamsplitter}
 		B_{ij}(T_S) = \begin{pmatrix}
 			\sqrt{T_S} \,\mathbb{1}& \sqrt{1-T_S} \,\mathbb{1} \\
 			-\sqrt{1-T_S} \,\mathbb{1}& \sqrt{T_S} \,\mathbb{1}
 		\end{pmatrix},
 	\end{equation}
 	where $T_S$ is the beam splitter transmissivity.
 	
 	The beam splitter acts on the quantum states of the two
 	mode field via an infinite-dimensional unitary operator
 	corresponding 
 	of the above action given by~\cite{arvind-1994,arvind1995,arvind-1995}
 	\begin{equation}\label{beam}
 		\mathcal{U}  (B_{ij}(T_S)) = \exp[ \theta
 		(\hat{a}_i^{\dagger} \hat{a}_j-\hat{a}_i
 		\hat{a}_j^{\dagger})],
 	\end{equation}
 	where $\theta$ is related to the transmissivity via the
 	relation $T_S = \cos^2 \theta$.

 	\par
 	\noindent{\bf  Two mode squeezer\,:}
 	It acts on the quadrature operators 
 	$(\hat{q}_{i}$, $\hat{p}_{i}$, $\hat{q}_{j}$, $\hat{p}_{j})^T$ through the matrix
 	\begin{equation}\label{eq:tms}
 		S_{ij}(r) = \begin{pmatrix}
 			\cosh r \,\mathbb{1}& \sinh r \,\mathbb{Z} \\
 			\sinh r \,\mathbb{Z}& \cosh r \,\mathbb{1}
 		\end{pmatrix},
 	\end{equation}
 	where $\mathbb{Z} = \text{diag}(1,\, -1)$.
 	The  infinite-dimensional unitary operator for the
 	two-mode   squeezer acting on the Hilbert space is given by
 	\begin{equation}\label{twomodesq}
 		\mathcal{U}  (S_{ij}(r)) = \exp[r(\hat{a}_i^{\dagger} \hat{a}_j^{\dagger}-\hat{a}_i
 		\hat{a}_j)].
 	\end{equation}

 	\subsection{Wigner characteristic function}
 	The Wigner characteristic function  for a state of
 	an $n$-mode quantum system represented by the density
 	operator $\hat{\rho}$  can be written as
 	\begin{equation}\label{wigdef}
 		\chi(\Lambda) = \text{Tr}[\hat{\rho} \, \exp(-i \Lambda^T
 		\Omega \,\hat{\xi})],
 	\end{equation}
 	where $\hat{\xi} = (\hat{q_1}, \hat{p_1},\dots
 	\hat{q_n}, \hat{p_n})^T$,  $\Lambda = (\Lambda_1, \Lambda_2,
 	\dots \Lambda_n)^T$ with  $\Lambda_i = (\tau_i, \sigma_i)^T
 	\in \mathcal{R}^2$.
 	Using the above equation, the Wigner characteristic
 	function for a single mode Fock state $|n\rangle$ evaluates to
 	\begin{equation}\label{charfock}
 		\chi_{|n\rangle}(\tau,\sigma)=\exp  \left[-
 		\frac{\tau^2}{2}-\frac{\sigma^2}{2} \right]\,L_{n}\left(
 		\tau^2 + \sigma^2 \right),
 	\end{equation}
 	where $L_n(x)$ is the Laguerre polynomial.
 	
 	First-order moments for an $n$-mode  quantum system  can be
 	written as
 	\begin{equation}
 		\bar{\hat{\xi}} = \langle  \hat{\xi } \rangle =
 		\text{Tr}[\hat{\rho} \,\hat{\xi}],
 	\end{equation}
 	which we call the displacement vector.  Similarly,
 	second-order moments can be conveniently arranged in the
 	form of a real symmetric $2n\times2n$ matrix,
 	known as the covariance matrix given by
 	\begin{equation}\label{eq:cov}
 		V = (V_{ij})=\frac{1}{2}\langle \{\Delta \hat{\xi}_i,\Delta
 		\hat{\xi}_j\} \rangle,
 	\end{equation}
 	where $\Delta \hat{\xi}_i = \hat{\xi}_i-\langle \hat{\xi}_i
 	\rangle$, and $\{\,, \, \}$ denotes the anti-commutator operation.
 	
 	Gaussian states, which play a very important role in
 	quantum optics and various QIP schemes are defined as states
 	with a Gaussian Wigner function or Gaussian Wigner characteristic function.
 	Such states are
 	completely specified by their displacement vectors and
 	covariance matrices. For a Gaussian state with displacement
 	vector $\bar{\hat{\xi}}$ and  covariance matrix $V$ the Wigner
 	characteristic function takes a simple
 	form~\cite{Weedbrook-rmp-2012, olivares-2012}:
 	\begin{equation}
 		\label{wigc} \chi(\Lambda)
 		=\exp[-\frac{1}{2}\Lambda^T (\Omega V \Omega^T) \Lambda- i
 		(\Omega \,\bar{\hat{\xi}} )^T\Lambda].  
 	\end{equation}
 	Therefore, the Wigner characteristic function of a
 	single mode coherent state with displacement
 	$\bar{\hat{\xi}}=(d_q,d_p)^T$   evaluates to
 	\begin{equation}
 		\chi_\text{coh}(\Lambda)= \exp \left[-\frac{1}{2}(\tau
 		^2+\sigma ^2)-i (\tau  d_p-\sigma  d_q)\right].
 	\end{equation}

 	Given a  symplectic transformation $S$  and  the
 	associated infinite dimensional unitary representation
 	$\mathcal{U}(S)$, the transformation of the  density
 	operator is given by 
 	$\hat{\rho} \rightarrow \,\mathcal{U}(S) \hat{\rho}
 	\,\mathcal{U}(S)^{\dagger}$.
 	Similarly,  the displacement vector, the covariance
 	matrix,  and the Wigner characteristic function transform as
 	by~\cite{arvind1995,olivares-2012,Weedbrook-rmp-2012}
 	\begin{equation}\label{transformation} 
 		\bar{\hat{\xi}}\rightarrow S \bar{\hat{\xi}},\quad
 		V\rightarrow SVS^T,\quad  \text{and} \,\,\chi(\Lambda)
 		\rightarrow \chi(S^{-1}\Lambda).
 	\end{equation}

 
 %
 
 \end{document}